\documentclass[aps,pre,12pt,a4paper]{revtex4-2}
\usepackage{epsfig}
\usepackage{graphicx}
\usepackage{amsmath,amssymb,color}
\usepackage[english]{babel}
\usepackage{natbib}
\usepackage{graphicx}
\usepackage{float}
\usepackage{physics}
\parskip=\medskipamount
\usepackage{subfigure}
\usepackage{comment}
\usepackage[table]{xcolor}
\usepackage{booktabs}        
\usepackage{array}           
\usepackage[capitalize,nameinlink,noabbrev]{cleveref}
\newcommand{\eq}[1]{(\ref{#1})}
\newcommand{\fig}[1]{Fig.~\ref{#1}}

\newcommand{\be}{\begin{equation}}
\newcommand{\ee}{\end{equation}}
\newcommand{\beq}{\begin{equation}}
\newcommand{\eeq}{\end{equation}}
\newcommand\disp{\displaystyle}

\newcommand{\la}{\left<}
\newcommand{\ra}{\right>}
\newcommand{\eps}{\varepsilon}

\begin{document}

\title{Probing phase transitions of finite directed polymers near a corrugated wall via two-replica analysis}
\author{Ruijie Xu$^1$, and Sergei Nechaev$^{2,1,3}$}
\affiliation{$^1$Beijing Institute of Mathematical Sciences and Applications (BIMSA), Yanqi Lake, Huairou District, Beijing 101408, China \\ $^2$LPTMS, CNRS -- Universit\'e Paris Saclay, 91405  Orsay Cedex, France \\ 
$^3$Laboratory of Complex Networks, Center for Neurophysics and Neuromorphic Technologies, Moscow, Russia}

\begin{abstract}

We study the pinning transition in a (1+1)-dimensional lattice model of a fluctuating interface interacting with a corrugated impenetrable wall. The interface is modeled as an $N$-step directed one-dimensional random walk on the half-line $x \ge 0$. Its interaction with the wall is described by a quenched, site-dependent, short-ranged random potential $u_j$ ($j = 1,\ldots,N$), distributed according to $Q(u_j)$ and localized at $x = 0$. By computing the first two disorder--averaged moments of the partition function, $\langle G_N \rangle$ and $\langle G_N^2 \rangle$, and by analyzing the analytic structure of the resulting expressions, we derive an explicit criterion for the coincidence or distinction of the pinning transitions in annealed and quenched systems. We show that, although the transition points of the annealed and quenched systems are always different in the thermodynamic limit, for finite systems there exists a “gray zone” in which this difference is hardly detectable. Our results may help reconcile conflicting views on whether quenched disorder is marginally relevant.

\end{abstract}

\maketitle

\section{Introduction}
\label{sect:1}

Wetting is one of the most extensively studied phenomena in the statistical physics of interfaces \cite{dietrich}. In general terms, wetting refers to the pinning of an interface by a solid, impenetrable substrate. The problem of wetting, more precisely, the pinning–depinning transition of an interface governed by its interaction with a corrugated impenetrable wall, has been addressed in numerous studies since the mid-1980s (see, for example, \cite{abraham1,abraham2,vanleew,chui_weeks} and references therein).

In 1986-1988, Forgacs et al. \cite{forgacs1,forgacs2} developed a perturbative renormalization group (RG) approach to (1+1)-dimensional wetting in a disordered potential. Around the same time, Grosberg and Shakhnovich \cite{gro} applied RG techniques to study the localization transition in ideal heteropolymer chains with quenched random chemical ("primary") structure at a point-like potential well in $D$-dimensional space. Many of their conclusions for $D=3$ align with those of \cite{forgacs1,forgacs2}. Both studies offered important insights into the thermodynamics near the transition from delocalized (depinned) to localized (pinned) regimes in the presence of quenched chemical disorder.

The case of $(1+1)D$ wetting in a potential with a bimodal periodic energy distribution was first examined in \cite{grosberg,nech_zhang}, and later explored in greater generality in subsequent works \cite{swain, burk, bauer, mont}. However, the key question -- What is the average transition temperature $T_{tr}$ for the pinning–depinning transition in a quenched random potential? -- remained still unanswered. Temperature enters into the problem through the Boltzmann weight $\beta_j = e^{u_j/T}$, where $u_j$ is the interaction energy between the $j$th segment of the fluctuating interface and the random substrate. The works \cite{forgacs1, forgacs2, gro} suggested that the average transition temperature $T_{tr}$ in the quenched random potential coincides with the annealed transition temperature $T_{an}$, defined for a system with preaveraged Boltzmann weights $\beta=\la\beta_j\ra= \la e^{u_j/T} \ra$.

In 1992, Derrida, Hakim, and Vannimenus \cite{derrida} revisited the $(1+1)D$ wetting model and, using RG techniques, showed that for a Gaussian distribution of the random potential, the disorder is marginally relevant. Subsequent studies by other authors \cite{stepanow,tang}, employing different approaches, arrived at the same conclusion.
These works initiated a longstanding debate on the relevance of disorder in low dimensions. 

Recently, question of whether quenched and annealed pinning transition points coincide has attracted renewed attention within rigorous statistical mechanics. In many probabilistic works authors have identified precise conditions under which quenched and annealed critical pinning transition points either coincide or are different. These criteria are typically formulated in terms of the tail behavior of the underlying renewal process and are proved using detailed comparison inequalities, rather than through explicit computations of the partition function \cite{AlexanderZygouras2009, BirknerSun2010}.

A rich and influential number of works has emerged from the probabilistic community focused on pinning of random walks in disordered media. Representative works by Giacomin, Lacoin, Toninelli, and Derrida developed rigorous bounds for disorder relevance using tools such as fractional moment estimates, coarse graining, and change-of-measure techniques \cite{DerridaGiacominLacoinToninelli2009, GiacominToninelli2009, GiacominLacoinToninelli2010}. In these works, the Harris criterion plays a central role: it provides a prediction for whether quenched disorder is relevant, irrelevant, or marginal. The probabilistic literature places the Harris criterion on a rigorous footing by identifying precise regimes in which quenched and annealed critical points coincide, are strictly separated, or differ only through marginal, typically logarithmic, corrections. These results are formulated in terms of universal features of the renewal ("fist return") probability and proved via comparison inequalities and probabilistic bounds. The language adopted throughout these works is formulated at a high level of generality, relying on rigorous inequalities, rather than on explicit computations of partition functions or a detailed analysis of the analytic structure of the free energy near the critical point for specific systems.

A particularly delicate question is the analysis of marginal relevance, where logarithmic corrections determine whether disorder shifts the critical threshold, but only asymptotically \cite{GiacominLacoinToninelli2010}. More recent contributions have extended this framework to universality in marginal regimes and continuum scaling limits \cite{CaravennaSunZygouras2017, Zygouras2024}.

Similarly, Alexander and Zygouras established rigorous results for quenched and annealed critical points in polymer pinning models, showing that the two coincide only under specific regimes of return exponent and temperature \cite{AlexanderZygouras2009}. Birkner and Sun obtained analogous results in random walk pinning, demonstrating strict inequalities between quenched and annealed critical values in various dimensions and underlying walk structures \cite{BirknerSun2010}. These probabilistic considerations emphasize "typical" structural results offering deeper generic insights than exact solutions for specific models of disordered systems.

At the same time, it is worth emphasizing that, in certain situations, the exact solution of specific models can play a complementary and meaningful role. While such solutions lack the generality of probabilistic approaches, they allow to probe fine, model-dependent features of the transition, including subtle finite-size effects and small quantitative differences that may remain invisible from a purely asymptotic or inequality-based perspective. In particular, exact or semi-exact analyses can reveal tiny shifts in critical behavior that are difficult to detect within general frameworks, yet may be relevant for numerical studies and concrete realizations of disordered systems.

So, in contrast to the rigorous probabilistic studies discussed above, the present work adopts a more explicit, moment-based approach tailored to a concrete lattice model of a directed walk near a corrugated wall with quenched site-dependent disorder. By computing the first two disorder-averaged moments of the partition function, $\langle G_N \rangle$ and $\langle G_N^2 \rangle$, we derive explicit conditions under which pinning transition points in quenched and annealed systems coincide or are distinct. We show that, in the thermodynamic limit, the transition points of annealed and quenched systems are always different, although for finite systems there exists a “gray zone” in which this distinction may be hardly noticeable. We find that for finite system the transition points depend strongly on the type of the disorder distribution $Q(u_j)$. Analyzing three representative types of disorder, Poissonian, bimodal, and Gaussian, we establish the criterion determining when the transition points in the quenched and annealed finite systems coincide and when they differ.

Our results thus could help reconcile opposing viewpoints on whether quenching is marginally relevant, by highlighting the role of finite-size effects and moment-based criteria in determining the effective transition point. From a practical standpoint, since numerical simulations and experimental studies typically deal with finite systems, it is worth investigating how close the transition points for quenched and annealed finite chains actually are, and how their difference scales with system size. Questions of this type, while not addressed by asymptotic probabilistic methods, may play a crucial role in interpreting numerical data and connecting rigorous results with observable phenomena. 

The two-replica approach considered in our work offers a valuable framework for studying disordered systems, particularly in finite-size models, by allowing for an exact treatment of disorder-averaged first and second moments of the partition function. One of its main advantages is that it captures correlations induced by the random environment between replicas, thereby providing a precise handle on the subtle differences between quenched and annealed behavior. Thus, for finite systems and numerical investigations, the two-replica approach offers a practical and theoretically controlled tool for quantitative probing disorder effects. Meanwhile, the two-replica approach has obvious limitation since it does not ensure direct access to the full probability distribution of the free energy. Moreover, while it highlights the existence of differences between quenched and annealed systems, it does not automatically yield detailed information on the analytic structure of the free energy near the critical point. 

The paper is structured as follows. In Section \ref{sect:2} we formulate the model, in Section \ref{sect:3} we compute the location of the transition point for averaged (one--replica) lattice partition function. In Section \ref{sect:4} we provide the analysis of analytic structure and transition point of the two--replica partition function on a lattice with a diagonal hopping in a wedge $(k\ge 1, m\ge 1)$ with different Boltzmann weights at the walls $(k=0,m>0)$, $(k>0,m=0)$ and at the corner $(k=0,m=0)$. In Section \ref{sect:5} we summarize obtained results. Some auxiliary results and computational details are presented in the Appendix.

\section{The model}
\label{sect:2}

Consider a directed $N$-step one-dimensional lattice random walk on a half--line $k\ge 0$, representing the height of the fluctuating interface in a presence of a nonhomogeneous impermeable boundary located at $k=0$. The interaction of the random walk with the boundary is described by the random site-dependent potential $u_j$ ($j=1,...,N$). The partition function $G_N(k)$ of $N$-step paths starting at point $k=0$ and ending at point $k\ge 0$ satisfies the "backward $\bullet\hspace{-4pt}\leftarrow$ master equation" where $\bullet$ designates the point $k=0$ where the potential is located and $\leftarrow, \rightarrow$ are "backward" and "forward" steps of random walk on the 1D lattice: 
\be
\begin{cases} G_{j+1}(k) = G_j(k-1) + G_j(k+1) + \left(e^{-u_{j+1}/T}-1\right) \delta_{k,0} G_N(k+1)  \medskip \\
G_j(k) = 0 & \quad k<0 \medskip \\
G_{j=0}(k) = \delta_{k,0} &
\end{cases}
\label{eq:01}
\ee
In \eq{eq:01}, $\delta_{k,0}$ is the Kronecker delta-function. To shorten notations we set $T=1$ supposing that the potential $u_j$ is dimensionless. In what follows we consider three types of the disorder distribution of $u_j$: 
\begin{enumerate}
\item The Poissonian distribution:
\be
Q(u_j=r) = \frac{\mu^{r}e^{-\mu}}{r!}
\label{eq:02}
\ee
where $\mathbb{E} (u_j) = \mu$;
\item The Asymmetric bimodal distribution with values $\{+u, -u\}$ chosen with probabilities $p$ and $1-p$ ($0\le p\le 1$):
\be
Q(u_j) = p \delta_{u_j, +u} + (1-p) \delta_{u_j, -u}
\label{eq:03}
\ee
where $\mathbb{E}(u_j) = u(2p-1)$;
\item The Gaussian distribution with nonzero mean:
\be
Q(u_j=\rho) = \frac{1}{\sqrt{2\pi\sigma^2}} e^{-\frac{(\rho-\nu)^2}{2\sigma^2}} 
\label{eq:04}
\ee
where $\mathbb{E}(u_j) = \nu$ and $\mathbb{E}\left((u_j-\nu)^2\right)=\sigma^2$.
\end{enumerate}
Let us emphasize that our consideration is quite general and is not restricted with only these specific types of a substrate disorder.

The question addressed in this work is as follows. Increasing local attraction of the random walk to the boundary at $k=0$, we force the random walk to pin (localize) at the axis $k=0$. The "pinning" transition from delocalized to localized regimes is manifested in the asymptotic behavior of the entropy $S_N(k)=-\ln G_N(k)$ at $N\gg 1$ and happens critically:
\be
\lim_{N\to\infty} \frac{\ln G_N(k)}{N} =
\begin{cases}
\mathrm{const} & \mbox{below $\beta_{cr}$} \medskip \\
\phi\left(\beta_1,...,\beta_N\right) > 0 & \mbox{above $\beta_{cr}$}
\end{cases}
\label{eq:05}
\ee
where $\beta_{cr}$ depends on the distribution $Q(u_j)$, while $\phi\left(\beta_1,...,\beta_N \right)$ is in genera a function of all $\beta_j$ ($j=1.,,,N$) -- for a concrete example see Eqs. \eq{eq:b<2}--\eq{eq:b>2} below. 

In the study of the pinning transition in an averaged random potential, referred as the annealed model, the random distribution of Boltzmann weights ${\beta_j}$ is replaced by their values averaged over the disorder distribution $Q(u_j)$, namely,
$$
\beta = \la e^{u_j}\ra_{Q(u_j)}
$$
for all $j = 1, \dots, N$. This averaging produces a homogeneous system without randomness at the boundary $k = 0$, which is known to undergo a transition at a critical value $\beta_{\mathrm{cr}}^{(a)}$ of the annealed partition function $\langle G_N(\beta_1, \dots, \beta_N ,|, x) \rangle$.

Turning to the pinning in the "quenched" model, define the disorder-dependent entropy $S_N({\beta_1,...,\beta_N}|k) = -\ln G_N({\beta_1,..., \beta_N}|k)$ for a specific ("quenched") sequence ${\beta_1, ..., \beta_N}$, and compute the average $\la S_N({\beta_1,...,\beta_N}|k) \ra_{Q(u_j)}$ over the distribution $Q(u_j)$ for all $j=1,...,N$. We then examine the critical behavior, which is expected to occur now at $\beta_{cr}^{(q)}$, corresponding to averaging of the logarithm of the partition function.

We are interested in the question whether the transition point, $\beta_{cr}$, in models with annealed and quenched disorders coincide or are distinct. Addressing this question requires averaging the logarithm of the partition function over the disorder -- a notoriously difficult problem. A commonly used approach is the replica trick \cite{edwards}, which involves averaging the $n$-th power of the partition function, $\la G_N^n \ra$, over the distribution $Q(u_j)$, and then extracting the average entropy using the relation $\la S_N \ra = \lim_{n \to 0} \frac{\la G_N^n \ra - 1}{n}$.  However, if we are solely interested in whether the pinning transition points coincide in annealed and quenched models, the problem admits significant simplification. In this case, it is sufficient to compute the transition points of the first two disorder-averaged moments, $\la G_N \ra$ and $\la G_N^2 \ra$, in the thermodynamic limit $N\to\infty$. These computations provide the answer whether the transition points differ in annealed and quenched disorder cases.  

\section{One--replica lattice partition function on a half--line}
\label{sect:3}

\begin{figure}[ht]
\includegraphics[width=0.9\textwidth]{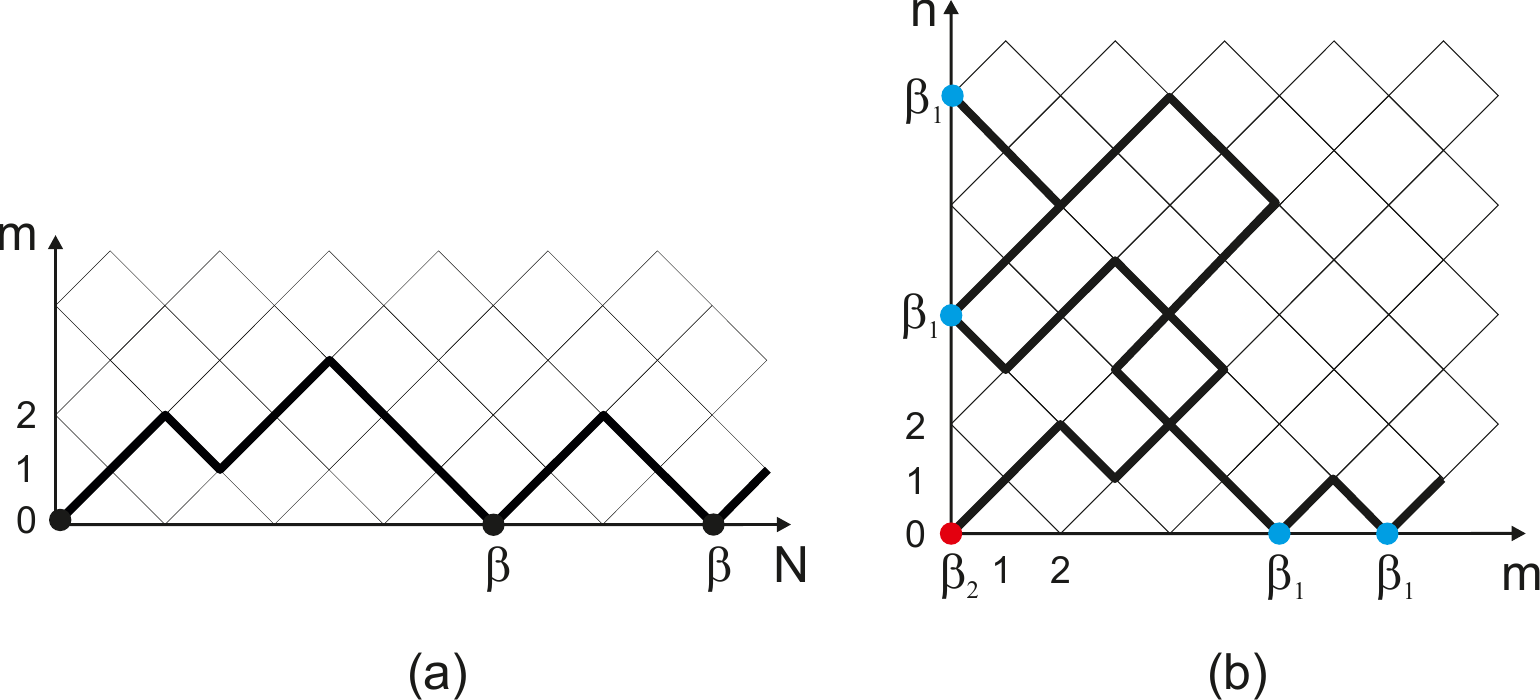}
\caption{Random walk interacting with the sticky boundary: (a) one-replica model with the uniform weight $\beta=\la e^{u_j} \ra_{Q(u_j)}$ along the boundary $k=0$; (b) two-replica model with two different weights: $\beta_1$ (shown in blue) along the boundaries $k=0$ and $m=0$ (except the corner) and $\beta_2$ (shown in red) at the corner $(k,m)=(0,0)$.}
\label{fig:01}
\end{figure}

The partition function $Z_N(k) = \la G_N(k) \ra$ averaged over the distribution of the disorder in the site-dependent potential $u_j$ can be computed directly by averaging l.h.s. and r.h.s. of \eq{eq:01}. Taking into account that $u_{j+1}$ and $G_j(k)$ are independent, we may write:
\be
\begin{cases}
Z_{j+1}(k) = Z_j(k-1) + Z_j(k+1) + (\beta-1)\delta_{k,0} Z_j(k+1) & \quad k\ge 0 \medskip \\
Z_j(k) = 0 & \quad k<0 \medskip \\
Z_{j=0}(k) = \delta_{k,0} &
\end{cases}
\label{eq:06}
\ee
where $\beta \equiv \la \beta_{j} \ra = \la e^{u_j} \ra_{Q(u_j)}$. In \fig{fig:01}a we schematically depict the one--replica model on a diagonal lattice with uniform weights $\beta$ at the boundary $m=0$, while \fig{fig:01}b illustrates the two-replica model with different values $\beta_1$ at boundaries $m=0$ and $n=0$ (except the corner) and $\beta_2$ at the corner $(m,n)=(0,0)$ --- this case is the subject of the discussion in Section \ref{sect:4}.

We solve \eq{eq:06} using the generating functions method. Define
\be
R(k,t) = \sum_{j=0}^{\infty} Z_j(k) t^j; \quad W(x,t) = \sum_{k=0}^{\infty} R(k,t) x^k
\label{eq:07}
\ee
Rewrite \eq{eq:06} in terms of generating functions $R(k,t)$ and $W(x,t)$:
\be
t^{-1} W(x,t) = t^{-1} + (x+x^{-1}) W(x,t) - x^{-1}R(0,t) + (\beta-1) R(1,t)
\label{eq:08}
\ee
At $k=0$ Eq.\eq{eq:06} reads $Z_{j+1}(0) = \beta Z_j(1)$ which for generating function $R(k,t)$ gives
\be
t^{-1} R(0,t) = t^{-1} + \beta R(1,t)
\label{eq:09}
\ee
Expressing $R(1,t)$ in terms of $R(0,t)$ and substituting \eq{eq:09} back into \eq{eq:08}, we arrive at the following equation
\be
\left(1 - t(x+x^{-1})\right) W(x,t) = \beta^{-1} + (1-\beta^{-1} - t x^{-1}) R(0,t)
\label{eq:10}
\ee
Let $K(x)=1 - t(x+x^{-1})$ be the kernel of \eq{eq:10}. The solutions of equation $K(x)=0$ are
\be
x_{\pm}=\frac{1\pm \sqrt{1-4t^2}}{2t}
\label{eq:11}
\ee
The branch $x_{-}$ satisfies the formal series expansion in $t$. Substituting $x_0\equiv x_{-}$ into \eq{eq:10} we get the equation for $R(0,t)$:
\be
0 = \beta^{-1} - (t x_0^{-1}-1+\beta^{-1}) R(0,t)
\label{eq:12}
\ee
The substitution is valid due to the composition law for multivariate formal power series \cite{2013Formal}. Thus, we have:
\be
R(0,t) = \frac{x_0}{t\beta -(\beta-1)x_0} = \frac{1-\sqrt{1-4t^2}}{2t^2\beta-(\beta-1)\left(1- \sqrt{1-4t^2}\right)}
\label{eq:13}
\ee
At the localization transition point the generating function $R(0,t)$ diverges and the transition point itself is determined by the equation nullifying the denominator of \eq{eq:13}:
\be
t=\frac{\sqrt{\beta-1}}{\beta}
\label{eq:14}
\ee
In the thermodynamic limit the critical value $t=t_{cr}$ is defined by the equation $\sqrt{1-4t_{cr}^2}=0$, thus giving $t_{tr} = \tfrac{1}{2}$, which being substituted into \eq{eq:14}, provides $\beta_{tr}=2$.

\section{Two--replica lattice partition function in a quarter of a plane}
\label{sect:4}

\subsection{"Boundary" vs "corner" localization in first quadrant on a diagonal lattice}
\label{sect:4a}

The equation on two-replica partition function, $W_N = \la G_N^2 \ra$, averaged over the distribution of $u_j$ can be obtained by averaging square of l.h.s. and r.h.s. of \eq{eq:01} with correct adjustment of boundary weights. Taking two copies, $G_j(m)$ and $G_j(n)$ with the same distribution of disorder, $u_j$, and denoting 
$$
W_j(m,n) = \la G_j(m) G_j(n) \ra_{Q(u_j)},
$$ 
we can easily derive the equation for the partition function of the ensemble of paths on a square diagonal lattice in a first quarter of the plane $(m,n)$. Schematically the corresponding 2D system is depicted in \fig{fig:01}b with two different weights: $\beta_1$ at boundaries (except the corner), and $\beta_2$ at the corner $(m,n)=(0,0)$. The difference between annealed and quenched systems is reflected in weights $\beta_1$ and $\beta_2$: 
\be
\left\{
\begin{array}{l}
\mbox{$\beta_1 = \la e^{u_j} \ra_{Q(u_j)}\equiv \beta$ and $\beta_2 = \la e^{u_j} \ra_{Q(u_j)}^2 = \beta^2$ for annealed system} \medskip \\
\mbox{$\beta_1 = \la e^{u_j} \ra_{Q(u_j)}\equiv \beta$ and $\beta_2 = \la e^{2u_j} \ra_{Q(u_j)} = \alpha \neq \beta^2$ for quenched system}
\end{array} \right.
\label{eq:ann-qu}
\ee 
Taking into account (as in 1-replica case) the independence of $u_{j+1}$ and $G_j^2$, we arrive at the following equation for a two-replica partition function $W_j(m,n)$:
\be
\left\{
\begin{array}{l}
W_{j+1}(m,n) = W_j(m-1,n-1) + W_j(m-1,n+1)+ W_j(m+1,n-1)+ W_j(m+1,n+1) + \medskip \\
\hspace{3cm} (\beta-1)\delta_{m,0} \left(W_j(m+1,n-1) + W_j(m+1,n+1) \right) + \medskip \\
\hspace{3cm} (\beta-1)\delta_{n,0} \left(W_j(m-1,n+1) + W_j(m+1,n+1) \right) + \medskip \\
\hspace{3cm} (\alpha - 2\beta+1)\delta_{m,0}\delta_{n,0} W_j(m+1,n+1) \hspace{2cm}  (m\ge 0, n\ge 0)\medskip \\
W_j(m,n) = 0 \hspace{9.3cm}  (m<0, n<0) \medskip \\
W_{j=0}(m,n) = \delta_{m,0}\delta_{n,0} 
\end{array}\right.
\label{eq:15}
\ee
Since generically $\alpha \neq \beta^2$, the localization of paths in two-replica system could behave differently than in one-replica (i.e. annealed) case. Let us provide the physical arguments behind this difference. In 2D system shown in \fig{fig:01}b, the localization of path emerges:
\begin{itemize}
\item[-] either due to adsorption of trajectories at the \emph{boundaries} of the first quadrant, and is controlled by the average weight $\beta$ at the boundaries, 
\item[-] or due to adsorption of trajectories at the \emph{corner} $(m,n)=(0,0)$, and is controlled by the weight $\alpha$ at the corner.
\end{itemize}
The corresponding values of $\beta$ and $\alpha$ for distributions of the disorders given by Eqs. \eq{eq:02}--\eq{eq:04},  are summarized in the Table \ref{tab:01}.

\begin{table}[ht]
\centering
\begin{tabular}{|c|c|c|}
\hline \hline
~Poissonian disorder Eq.\eq{eq:02}~ & ~Asymmetric bimodal disorder Eq.\eq{eq:03}~ & ~Gaussian disorder Eq.\eq{eq:04}~ \\ \hline  
$\beta = \sum_{r=0}^{\infty} e^r Q(r) = e^{(e-1)\mu}$ & $\beta = pe^{u}+(1-p)e^{-u}$ & $\beta = \int_{-\infty}^{\infty} e^{\rho} Q(\rho) d\rho= e^{\frac{\sigma^2}{2}+\nu}$ \\ 
$ ~\alpha = \sum_{r=0}^{\infty} e^{2r} Q(r) = e^{(e^2-1)\mu}$~ & $\alpha = p e^{2u}+(1-p)e^{-2u}$ & $~\alpha = \int_{-\infty}^{\infty} e^{2\rho} Q(\rho) d\rho = e^{2(\sigma^2+\nu)}$~ \\
\hline \hline
\end{tabular}
\caption{Weights $\beta$ and $\alpha$ for different disorders.}
\label{tab:01}
\end{table}

By varying the parameters of the distributions ($\mu$ for Poissonian, $(u, p)$ for asymmetric bimodal, and $(\nu, \sigma)$ for Gaussian), the weights $\beta$ and $\alpha$ are altered accordingly. Depending on the type of disorder, increasing these parameters could in principle lead to two distinct scenarios:
\begin{itemize}
\item[(i)] The critical value $\beta_{cr}$, associated with the \emph{boundary localization}, is reached first, while $\alpha$ remains below its critical threshold $\alpha_{cr}$ for corner localization; in this case the transition points for annealed and quenched disorder coincide and the transition occurs at the annealed value $\beta_{cr}$ in all replicas;
\item[(ii)] The critical value $\alpha_{cr}$ is reached first, indicating the \emph{corner localization}, while $\beta$ is still below $\beta_{cr}$; in this case transition points differ, indicating that the disorder is marginally relevant and the transition in the two-replica case occurs at $\alpha_{cr}$.
\end{itemize}

\subsection{Exact formal solution of a two-replica problem for arbitrary $\beta$ and $\alpha$} 
\label{sect:4b}

We solve the "backward" master equation \eq{eq:15} and determine the localization transition in the phase space $(\beta, \alpha)$ using the generating function method \cite{bousquet2010walks}. Define the following generating functions (compare with \eq{eq:06}):
\be
\begin{array}{ll}
\disp P(m,n,t) = \sum_{j=0}^{\infty} W_j(m,n) t^j; & \disp \quad Q(x,y,t) = \sum_{m=0}^{\infty} \sum_{n=0}^{\infty}P(m,n,t) x^m y^n; \medskip \\ \disp R_1(x,n,t) = \sum_{m=0}^{\infty} P(m,n,t) x^m;  & \disp \quad 
R_2(m,y,t) = \sum_{n=0}^{\infty} P(m,n,t) y^n \end{array}
\label{eq:16}
\ee
Writing equations for generating functions $Q(x,y,t), P(m,n,t), R_1(x,n,t), R_2(m,y,t)$, we get:
\be
\left\{\begin{array}{ll}
t^{-1}Q(x,y,t) = & t^{-1}+(x+x^{-1})(y+y^{-1})Q(x,y,t)-x^{-1}(y+y^{-1})R_2(0,y,t)- \medskip \\ & y^{-1}(x+x^{-1})R_1(x,0,t) +(\beta-1)\left((y+y^{-1})R_2(1,y,t)-y^{-1}P(1,0,t) \right) +\medskip \\ & (\beta-1)\left((x+x^{-1})R_1(x,1,t)-x^{-1}P(0,1,t) \right) + (xy)^{-1} P(0,0,t) + \medskip \\ & (\alpha-2\beta+1)P(1,1,t) \medskip \\
t^{-1}P(0,0,t) = & t^{-1}+\alpha P(1,1,t) \medskip \\
t^{-1}R_1(x,0,t) = & t^{-1}+\beta\left((x+x^{-1})R_1(x,1,t)-x^{-1}P(0,1,t) \right)+(\alpha-\beta)P(1,1,t) \medskip \\
t^{-1}R_2(0,y,t) = & t^{-1}+\beta\left((y+y^{-1})R_2(1,y,t)-y^{-1}P(1,0,t) \right)+(\alpha-\beta)P(1,1,t)
\label{eq:17}
\end{array}\right.
\ee
Denoting $\Delta_{xy}=(x+x^{-1})(y+y^{-1})$, $\Delta_x= x+x^{-1}$, $\Delta_y=y+y^{-1}$ and introducing two combinations
\be
c_1= \Delta_y R_2(1,y,t)-y^{-1}P(1,0,t); \quad c_2= \Delta_x R_1(x,1,t)-x^{-1}P(0,1,t),
\label{eq:18}
\ee
we rewrite \eq{eq:17} as follows:
\be
\left\{\begin{array}{ll}
t^{-1}Q(x,y,t) = & t^{-1}+\Delta_{xy}Q(x,y,t)-x^{-1}\Delta_y R_2(0,y,t)- y^{-1}\Delta_x R_1(x,0,t) - \medskip \\ &  (xy)^{-1} P(0,0,t) + (\beta-1)(c_1+c_2) + (\alpha-2\beta+1)P(1,1,t) \medskip \\
t^{-1}P(0,0,t) = & t^{-1}+\alpha P(1,1,t) \medskip \\
t^{-1}R_1(x,0,t) = & t^{-1}+\beta c_2+(\alpha-\beta)P(1,1,t) \medskip \\
t^{-1}R_2(0,y,t) = & t^{-1}+\beta c_1+(\alpha-\beta)P(1,1,t)
\label{eq:19}
\end{array}\right.
\ee
Extracting the combination $c_1 + c_2$ from the last two equations in \eq{eq:19}, substituting it into the first line of \eq{eq:19}, and then rewriting $P(1,1,t)$ in terms of $P(0,0,t)$, we obtain a single "backward" equation for the generating functions, suitable for the forthcoming analysis:
\begin{multline}
(1 - t\Delta_{xy})Q(x,y,t) = \frac{1}{\alpha} + \left(\frac{2}{\beta} - \frac{1}{\alpha} - 1 +  \frac{t}{x y}\right)P(0,0,t) + \\ \left(1-\frac{1}{\beta} - \frac{t(1 + x^2)}{x y}\right)R_1(x,0,t) + \left(1-\frac{1}{\beta} - \frac{t(1 + y^2)}{x y}\right)R_2(0,y,t) 
\label{eq:20}
\end{multline}
All generating functions are invariant with respect to the group of transformations generated by the substitutions $x\leftrightarrow x^{-1}$ and $y\leftrightarrow y^{-1}$. 

Consider the solutions of algebraic equation $1 - t\Delta_{xy}=0$:
\be
y_{\pm}=\frac{x \pm \sqrt{x^2 - 4 t^2 (1 + x^2)^2}}{2 t (1 + x^2)}
\label{eq:21}
\ee
Only the root $y_0\equiv y_{-}$ satisfies the formal series expansion of $t$. Substitute $y_0$ into \eq{eq:20} and into the same equation obtained under the transformation $x\to x^{-1}$. Denote for brevity: 
\be 
P\equiv P(0,0,t), \quad R_1(x)\equiv R_1(x,0,t), \quad R_2\equiv R_2(0,y,t)
\label{eq:def}
\ee
We get:
\be
\begin{array}{l}
\disp \frac{1}{\alpha} + \left(\frac{2}{\beta} - \frac{1}{\alpha} - 1 +  \frac{t}{x y_0}\right)P + \left(1-\frac{1}{\beta} - \frac{t(1 + x^2)}{x y_0}\right)R_1(x) + \left(1-\frac{1}{\beta} - \frac{t(1 + y_0^2)}{x y_0}\right)R_2=0 \medskip \\
\disp \frac{1}{\alpha} + \left(\frac{2}{\beta} - \frac{1}{\alpha} - 1 +  \frac{t x}{y_0}\right)P + \left(1-\frac{1}{\beta} - \frac{t(1 + x^2)}{x y_0}\right)R_1\left(\frac{1}{x}\right) + \left(1-\frac{1}{\beta} - \frac{tx(1 + y_0^2)}{y_0}\right)R_2=0
\end{array}
\label{eq:22}
\ee
Now we can eliminate $R_2$ from Eqs.\eq{eq:22}: we express $R_2$ using second line of \eq{eq:22} and substitute it in the first line of \eq{eq:22}. We obtain:
\begin{multline}
-P \frac{\beta t x \left(x^2-1\right) y_0 \left(-\alpha +\beta + y_0^2 (\alpha  (\beta -2)+\beta )\right)}{\alpha  \left(\beta \left(t x^2+t-x y_0\right)+x y_0\right) \left(\beta  \left(t y_0^2+t-x y_0\right)+x y_0\right)}-R_1(x)\frac{x\left(\beta t x \left(y_0^2+1\right)-\beta  y_0+y_0\right)}{\beta \left(t y_0^2+t-x y_0\right) + x y_0}+ \\ R_1\left(\frac{1}{x}\right)+\frac{\beta ^2 t x \left(x^2-1\right) y_0 \left(y_0^2+1\right)}{\alpha  \left(\beta \left(t x^2+t-x y_0\right)+x y_0\right) \left(\beta  \left(t y_0^2+t-x y_0\right)+x y_0\right)} = 0
\label{eq:23}
\end{multline}
Since $y_0$ is a root of a quadratic polynomial, any rational function of $y_0$ can be written in a form $f_1(x,t) y_0 + f_2(x,t)$ where $f_1(x,t)$ and $f_2(x,t)$ are rational functions of $x$ and $t$ only. Applying this idea to \eqref{eq:23}, we rewrite it in the following form:
\be
C_c(x,t)= \frac{\left(-\beta +x^2+1\right)}{(\beta-1)x^2-1} R_1(x) + R_1\left(\frac{1}{x}\right)+C_p(x,t)P
\label{eq:24}
\ee
where the coefficients $C_c(x,t)$ and $C_p(x,t)$ are linear functions of $y_0$ with rational coefficients in $x,t$:
\be
\begin{array}{ll}
C_c(x,t)= & \disp \frac{\beta ^2 (x-1) x (x+1) \left(\beta  t x^2+\beta  t\right)}{\alpha  \left(\beta  x^2-x^2-1\right) \left(\beta ^2 t^2+\beta ^2 t^2 x^4+2 \beta ^2 t^2 x^2-\beta  x^2+x^2\right)}y_0 - \medskip \\
& \disp \disp \frac{\beta (x-1)}{\alpha  \left(\beta  x^2-x^2-1\right) \left(\beta ^2 t^2+\beta ^2 t^2 x^4+2 \beta ^2 t^2 x^2-\beta  x^2+x^2\right)} \medskip \\
C_{p}(x,t)= & \disp \frac{\beta  t x \left(x^4-1\right)\left(\alpha -\beta ^2\right)}{\alpha  \left((\beta -1) x^2-1\right) \left(\beta ^2 t^2+\beta ^2 t^2 x^4+x^2 \left(-\beta +2 \beta ^2 t^2+1\right)\right)}y_0 + \medskip \\
& \disp \frac{(\beta -1) \beta  (x-1) (x+1) \left(\alpha  \beta  t^2+\alpha  \beta  t^2 x^4+x^2 \left(\beta +\alpha  \left(2 \beta  t^2-1\right)\right)\right)}{\alpha  \left((\beta -1) x^2-1\right) \left(\beta ^2 t^2+\beta ^2 t^2 x^4+x^2 \left(-\beta +2 \beta ^2 t^2+1\right)\right)}
\end{array}
\label{eq:coefficients}
\ee

Equation \eq{eq:24} is solvable. Let us currently assume $|x^2|<|\frac{1}{\beta-1}|$. The denominator of the  coefficient in front of $R_1(x)$ in \eq{eq:24} can be written as $((\beta-1)x^2-1)=\sum_{j=0}((\beta-1)x^2)^j$. Thus, terms involving $R_1(x)$ contain only non-negative degrees of $x$, while terms involving $R_1(1/x)$ contain only non-positive degrees of $x$. Denote by $[x^0]$ the operator of extracting $x^0$ terms of a series in $x$. Applying $[x^0]$ to \eqref{eq:24}, and recalling the connections of $R_1(x)$ and $R_1(1/x)$ to $P$ which follow from the definition \eq{eq:16} and \eq{eq:def}, we get:
\be
[x^0]C_c(x,t)=(\beta-1)P+P+P[x^0]C_p(x,t)
\label{eq:26}
\ee
Equation \eq{eq:26} is linear in $P$ and thus $P$ formally can be expressed as follows:
\be
P= \frac{[x^0]C_c(x,t)}{\beta+[x^0]C_p(x,t)}
\label{eq:26a}
\ee
Theoretically, the coefficients are computable since $[x^0]$ is equivalent to the contour integral $\frac{1}{2\pi i}\oint\frac{dx}{x}$. However, practically, $C_c(x,t)$ and $C_p(x,t)$ are complicated elliptic integrals and we are therefore led to consider an indirect method to simplify the resulting expressions.

\subsection{Simplification by the first return}
\label{sect:4c}

Note that the first term of $C_{p}(x,t)$ in \eqref{eq:coefficients} includes the factor $(\alpha - \beta^2)y_0$. When $\alpha = \beta^2$, this factor vanishes, and $C_p(x,t)$ becomes a rational function in $x$ and $t$. This special case is significantly simpler to analyze. Therefore, we begin by considering the case $\alpha = \beta^2$, and later extend the discussion to the general case using insights from this simpler setting.

To proceed, it is helpful to adopt a combinatorial approach. For this purpose, we define two auxiliary functions, $A(0,0)$ and $F(0,0)$:
\begin{itemize}
\item $A(0,0)$ is the generating function for paths that start at $(0,0)$ and return to $(0,0)$ for the first time at their final step. Importantly, we \emph{do not assign a weight} $\beta$ to this final step. Thus, $A(0,0)$ corresponds to the grand canonical partition function for "first return" paths.

\item $F(0,0)$ is the generating function for paths that start and end at $(0,0)$, where the interaction at the endpoint $(0,0)$ \emph{is weighted with} $\beta^2$. All other interactions along the path are assigned the usual weight $\beta$.
\end{itemize}

There exist a straightforward connections between the generating functions $A(0,0)$, $F(0,0)$, and $P$:
\be
F(0,0)=\frac{1}{1-\beta^2 A(0,0)}, \qquad P=\frac{1}{1-\alpha A(0,0)}
\label{eq:28}
\ee
The function $A(0,0)$ does not involve the interaction at $(0,0)$ and can be excluded from equations \eq{eq:28}. So, we arrive at the following expression
\be
P=\frac{\beta^2}{\beta^2-\alpha}\left(1-\frac{\alpha}{\alpha+(\beta^2-\alpha)F(0,0)}\right)
\label{eq:29}
\ee
The asymptotic behavior of $Q(0,0)$ depends on singularities of $F(0,0)$ and poles of the function $\alpha+(\beta^2-\alpha)F(0,0)=0$. Recall that $F(0,0)$ is the solution of \eqref{eq:26} with $\alpha=\beta^2$ and from the physical point of view this equation means that the quenched disorder is replaced by the annealed one.

\subsection{Asymptotic of $F(0,0)$ via bijection}
\label{sect:4d}

The 2D diagonal walk (DW) considered throughout our paper is the Cartesian product of two Dyck paths interacting via boundary terms. Thus, there exists a bijection \cite{beaton2018exact}:
$$
\mbox{DW from $(0,0)$ to $(0,0)$} \Longleftrightarrow \mbox{Pair of Dyck path $(Z_1, Z_2)$}
$$
The bijection is set explicitly as follows:
\be
\nearrow \leftrightarrow \{\nearrow,\nearrow\}; \quad
\swarrow \leftrightarrow \{\searrow,\searrow\} ; \quad
\nwarrow \leftrightarrow \{\searrow,\nearrow\}; \quad
\searrow \leftrightarrow \{\nearrow,\searrow\}
\ee
The map is injective by definition. Denote by $\#$ the number of steps of specific type in a path. The map is well-defined because, for any $n$-step path confined to the first quadrant, the condition
$$
(\#\nearrow) + (\#\searrow) - (\#\nwarrow) - (\#\swarrow) \geq 0
$$
for every $k$-step sub-path (with $k \leq n$) ensures that the walk $Z_1$ stays above the $x$-axis, and vice versa. Similarly, the condition
$$
(\#\nearrow) + (\#\nwarrow) - (\#\searrow) - (\#\swarrow) \geq 0
$$
guarantees that the walk $Z_2$ stays above the $x$-axis, and vice versa. Therefore, the map is bijective. In terms of generating function, the number of configurations satisfies the following relation:
\be
[t^{2j}]F(0,0)=Z^2_{2j}(0)=\left([t^{2j}]R(0,t)\right)^2
\label{eq:30}
\ee
$Z_{2j}(0)$ and $R(0,t)$ are defined by \eqref{eq:06} and \eqref{eq:07} and by $[t^{2j}](...)$ we denote the operator which is equivalent to the contour integral $\frac{1}{2\pi i}\oint \tfrac{dx}{x^{2j+1}}$. This integral representation is also known as Lattice Green Function \cite{2010Lattice}. Thus, the phase diagram is exactly the same as for equation \eqref{eq:13}, yielding $t_{tr} = \tfrac{1}{2}$ and a transition point is given by the value $\beta_{tr} = 2$. The only difference lies in squaring of the partition function.

The explicit asymptotics of the partition function $Z_{2j}(0)$ in different regimes are as follows:
\begin{enumerate}
\item[(i)] For $\beta<2$, the asymptotic is determined by the branch-cut, 
\be
Z_{2j}(0)\sim [t^{2j}](1-4t^2)^{1/2}\sim\frac{2^{2j}j^{-3/2}}{\Gamma(-\tfrac{1}{2})} \quad \Rightarrow \quad [t^{2j}]F(0,0)\sim \frac{4^{2j}j^{-3}}{\Gamma^2(-\tfrac{1}{2})}
\label{eq:b<2}
\ee
\item[(ii)] For $\beta=2$, the asymptotic is determined by the branch-cut,
\be
Z_{2j}(0)\sim[t^{2j}](1-4t^2)^{-1/2}\sim\frac{2^{-2j}j^{-1/2}}{\Gamma(\tfrac{1}{2})} \quad \Rightarrow \quad
[t^{2j}]F(0,0)\sim \frac{4^{2j}j^{-1}}{\Gamma^2(\tfrac{1}{2})}
\label{eq:b=2}
\ee
\item[(iii)] For $\beta>2$, the asymptotic is determined by the pole,
\be
Z_{2j}(0)\sim [t^{2j}]\left(\frac{\beta-1}{\beta^2}-t^2\right)^{-1}\sim \left(\frac{\beta^2}{\beta-1}\right)^j
\quad \Rightarrow \quad[t^{2j}]F(0,0)\sim \left(\frac{\beta^2}{\beta-1}\right)^{2j}
\label{eq:b>2}
\ee
\end{enumerate}

\subsection{Asymptotic behavior of $Q(0,0)$ for $1<\beta <2$}
\label{sect:4e}

Consider $\alpha$ and $\beta$ introduced in \eq{eq:ann-qu} as formal parameters that can take any positive values such that $\alpha \ge \beta^2$, since for any distribution of the disorder the condition $\alpha \ge \beta^2$ holds. By \eqref{eq:29}, the singularities of the generating function $P(0,0)$ arise from two sources: (a) the singularities of $F(0,0)$, and (b) the roots of $F(0,0)=\frac{\alpha}{\alpha-\beta^2}$. Since $F(0,0)$ is a formal series in $t$ with positive coefficients, for $\alpha > \beta^2$, case (b) gives rise to a new singularity in $P(0,0)$. 

To obtain the phase diagram of $P(0,0)$, we have no other ways, except estimating the value of $F(0,0)$. Our main focus is on the region $1<|\beta| < 2$, where the inequality $1/\sqrt{|\beta - 1|} > 1 > |x|$ holds, which is consistent with the domain of convergence for the definition of $R_1(x)$. For completeness, the case $\beta>2$ is discussed in Appendix \ref{app:1}. 

The operator $[x^0]$ is treated as taking the contour integral $C$, so we need to make sure $C$ does not cross the branch-cuts or singularities of any functions. We choose $|C|=1-\epsilon$. Since $1/\sqrt{|\beta-1|}>1>|x|$ and $|t|<1/4$, the contour $C$ does not cross the branch-cut of $R_1(x)$ or $y_0$. The consistency of $R_1(1/x)$ is due to the analytic continuation -- see, for example, \cite{fayolle1999random}. Eq.\eqref{eq:24} can be considered as the definition of $R_1(1/x)$ through $R_1(x)$ and $y_0$. Then, $R_1(1/x)$ as a function defined by $R_1(x)$ and $y_0$ is also well defined and compatible with the definition of $|C|$.

By \eqref{eq:26}, we have, 
\begin{multline}
F(0,0)=[x^0]\left(-\frac{(x-1) x (x+1) \left(\beta  \sqrt{x^2-4 \left(-t x^2-t\right)^2}+\beta  x-2 x\right)}{2 \left(\beta  x^2-x^2-1\right) \left(\beta ^2 t^2+\beta ^2 t^2 x^4+2 \beta ^2 t^2 x^2-\beta  x^2+x^2\right)}\right)= \\ - \frac{1}{2\pi i}\oint_C\frac{(x-1) (x+1) \left(\beta  \sqrt{x^2-4 \left(-t x^2-t\right)^2}+\beta  x-2 x\right)}{2 \left(\beta  x^2-x^2-1\right) \left(\beta ^2 t^2+\beta ^2 t^2 x^4+2 \beta ^2 t^2 x^2-\beta  x^2+x^2\right)}dx
\label{eq:31}
\end{multline}
where the contour $C$ surrounds the origin $x=0$ of the complex plain $x$ and $|C|=1-\eps$ ($\eps \to 0$). Since $F(0,0)$ is monotonous of $t$, the criticality is reached at $t\to t_{cr}= \tfrac{1}{4}$.

We may consider two insights of this problem. From a combinatorial point of view, we are working with a formal series in $t$. So in practice, we proceed as follows: (i) expand the function as a formal series in $t$, (ii) expand each term in this formal series as a series in $x$, (iii) collect the $[x^0]$--degree terms in this series and take the limit $t\to \tfrac{1}{4}$. 

Following this prescription, we choose the precision of our calculation to be $O^r(t)$ with $r=20$ and get the following approximate results for pairs $(\beta, \alpha_{cr})$ at which the long (though finite) path localizes:
\begin{table}[ht]
\centering
\begin{tabular}{cc}
\hline \hline
Value of $\beta$~ & ~Critical value $\alpha_{cr}$ for finite paths \\ \hline
\rowcolor{gray!20} $\beta=1.0$ & $\alpha_{cr}=11.9048$ \\ \hline
$\beta=1.2$ & $\alpha_{cr}=11.2764$ \\ \hline
\rowcolor{gray!20} $\beta=1.6$ & $\alpha_{cr}=9.98733$  \\ \hline
$\beta=1.8$ & $\alpha_{cr}=9.3802$ \\ \hline \hline
\end{tabular}
\caption{Critical values of $\alpha_{cr}$ at which chain localizes for a given value of $\beta$. The results are obtained with the precision $O^{20}(t)$.}
\label{tab:02}
\end{table}

Obtained values of $\alpha_{cr}$ are larger than the exact values corresponding to the transition point at $t\to\infty$. However higher precision in $t$ requires much heavier computation. To get a better result, consider the analytic insight and use the integral representation of \eqref{eq:31}.

The integral can be separated into two parts,
\[\frac{1}{2\pi i}\oint_C f(x)dx=\sum_{x\in\text{poles}} \Res(f(x))+\frac{1}{2\pi i}\oint\limits_{\rm branch~cut}f(x)dx = \Res(f) + \mathrm{Br}(f)\]

Consider first the poles inside the contour. A simple observation is as follows: if $f(x^2)$ is a function of $x^2$, then for any pole $x=X$, the value $x=-X$ is also a pole. Thus, we have:
\be
\frac{1}{2\pi i}\oint\frac{f(x^2)}{(x-X)(x+X)}dx=\frac{f(X^2)}{2X}+\frac{f((-X)^2)}{-2X}=0.
\ee
In \eqref{eq:31}, the parts with square root are the functions of $x^2$. Thus, they do not contributes to the residues and we get:
\be
\Res\left(F(0,0)\right)=\sum_{i}\Res\left(-\frac{(x-1)(x+1) \left(\beta  x-2 x\right)}{2 \left(\beta  x^2-x^2-1\right) \left(\beta ^2 t^2+\beta ^2 t^2 x^4+2 \beta ^2 t^2 x^2-\beta  x^2+x^2\right)}\right)\bigg|_{x=X^{(i)}}.
\label{eq:res}
\ee
The summation runs over all $X^{(i)}$ inside the unit circle. Note that the first factor $X_1(x)=(\beta-1)x^2-1$ in the denominator of \eq{eq:res} always has roots outside the unit circle, while the second factor can be rewritten as follows:
\be
X_2(x)=\left(\beta ^2 t^2+\beta ^2 t^2 x^4+2 \beta ^2 t^2 x^2-\beta  x^2+x^2\right)=x^2\left((1-\beta)+\beta^2t^2\left(x+\frac{1}{x}\right)^2\right)
\label{eq:2factor}
\ee
The roots of $X_2(x)$ is convenient to represent in polar coordinates using the substitution $x=r e^{i\theta}$. The roots of $X_2(x)$ are the solutions of the pair of equations: 
\be
\left(r+\frac{1}{r}\right)\cos\theta=\sqrt{\frac{\beta-1}{\beta^2t^2}}; \quad \left(r-\frac{1}{r}\right)\sin\theta=0.
\ee
Thus, we either have $r=1$, or $\theta=0$. Roots $X^{(i)}$ for $r=1$ lie outside of $C$. For $\theta=0$, since $\left|r+\frac{1}{r}\right|>2$, we have an extra condition $t<\frac{\sqrt{\beta-1}}{2\beta}$. This factor is a polynomial of $x^2$, and has two roots $\pm X^{(1,2)}$. The residue part finally reads,
\be
\Res(F(0,0))=\begin{cases}
\disp -2\Res\left(\frac{(x-1)(x+1) \left(\beta  x-2 x\right)}{2 X_1(x) X_2(x)}\right)\bigg|_{x=X_1} & \disp t<\frac{\sqrt{\beta-1}}{2\beta} \\
0 & \disp t\geq \frac{\sqrt{\beta-1}}{2\beta}
\end{cases}
\label{eq:36}
\ee
Further, since $x^2-1<0$ and $\beta-2<0$, the residue part is always negative. Recalling that we are trying to find the largest $t$ such that $F(0,0)=\frac{\alpha}{\alpha-\beta^2}$ is satisfied, we are taking the limit $t\to \tfrac{1}{4}$. Thus we always have $Res(F(0,0))=0$ for $\beta<2$.

The second contribution is from the integral on the branch-cut which in the thermodynamic limit $t\to \tfrac{1}{4}$ is $[-1,1]$. The terms without square root do not contribute. The argument of the function $\sqrt{x^2-4 \left(-t x^2-t\right)^2}$ differs by $\pi$ for the upper and lower limits, thus the contour integral becomes $2$ times the value of the integral of the one side:
\be
\mathrm{Br}(F(0,0))=-\frac{1}{\pi }\int_{-1}^1\frac{4\beta  (1-x^2)^2 }{\left((\beta -1) x^2-1\right) \left(\beta ^2+\beta ^2 x^4+2 \left(\beta ^2-8 \beta +8\right) x^2\right)}dx
\label{eq:33}
\ee
This equation permits us to express $F(0,0)$ as follows:
\be
F(0,0)=\mathrm{Br}(F(0,0))\label{eq:34}
\ee
and find the final expression for the critical corner weight $\alpha_{cr}$:
\be
\alpha_{cr} = \frac{\beta^2  F(0,0)}{F(0,0) - 1}
\label{eq:crit-alpha}
\ee
Eq.\eqref{eq:34} is an integral of a rational function and admits solutions for any $1 < \beta < 2$: for each $\beta$, we can find the corresponding critical value $\alpha_{cr}$. However, since Eq.\eqref{eq:34} involves quadratic roots, analytic computations become quite complicated. Therefore, in Table \ref{tab:03}, we provide numerical evaluations of the exact analytic expression for several representative pairs $(\beta, \alpha_{cr})$.

\begin{table}[ht]
\centering
\begin{tabular}{cc}
\hline \hline
Value of $\beta$~ & ~Exact critical value $\alpha_{cr}$ for infinite paths \\ \hline
$\beta=1.0$ & $\alpha_{cr}=\frac{4 (\pi-4)}{5\pi -16}=11.7575$ \\ \hline
\rowcolor{gray!20} $\beta=1.2$ & $\alpha_{cr}=11.0353$ \\ \hline
$\beta=1.6$ & $\alpha_{cr}=9.26537$ \\ \hline
\rowcolor{gray!20} $\beta=1.8$ & $\alpha_{cr}=8.02968$ \\ \hline
$\beta=1.999$ & $\alpha_{cr}=5.05185$ \\ \hline 
\rowcolor{gray!40} $\beta=\beta_{cr}=2$ & $\alpha_{cr}=4$ \\ \hline \hline 
\end{tabular}
\caption{Pairs of values ($\beta, \alpha_{cr}$) as in Table \ref{tab:02}, however here they are computed by numeric evaluation of integral \eq{eq:33} in the $t\to\tfrac{1}{4}$ limit.}
\label{tab:03}
\end{table}

The values of $\alpha, \beta_{cr}$ presented in Table \ref{tab:02} are quite remarkable. At the point $\beta = \beta_{cr} = 2$, which corresponds to the adsorption critical weight in the annealed system, the critical value of the corner weight $\alpha_{cr}$ reaches from above its exact boundary value $\alpha = 4$. Since, for any choice of parameters and for any disorder distribution $Q(u_j)$, the inequality $\alpha \ge \beta^2$ holds, quenched disorder is always relevant, and the quenched and annealed transition points are therefore different.  However, the point $\beta_{\mathrm{cr}} = 2$ is approached logarithmically, and as shown in Table \ref{tab:03}, for values of $\beta$ very close to but slightly less than 2, the critical corner weight $\alpha_{\mathrm{cr}}$ rapidly diverges from $\alpha_{\mathrm{cr}} = 4$. For example, at $\beta = 1.999$, $\alpha_{\mathrm{cr}}$ already reaches approximately 5.052. This behavior manifests the existence of a “gray zone” near $\beta_{\mathrm{cr}}$ that causes numerical ambiguities and leaves room for various speculations regarding the coincidence or difference of transition points in quenched and annealed systems.

\begin{figure}[ht]
\centering
\includegraphics[width=0.75\linewidth]{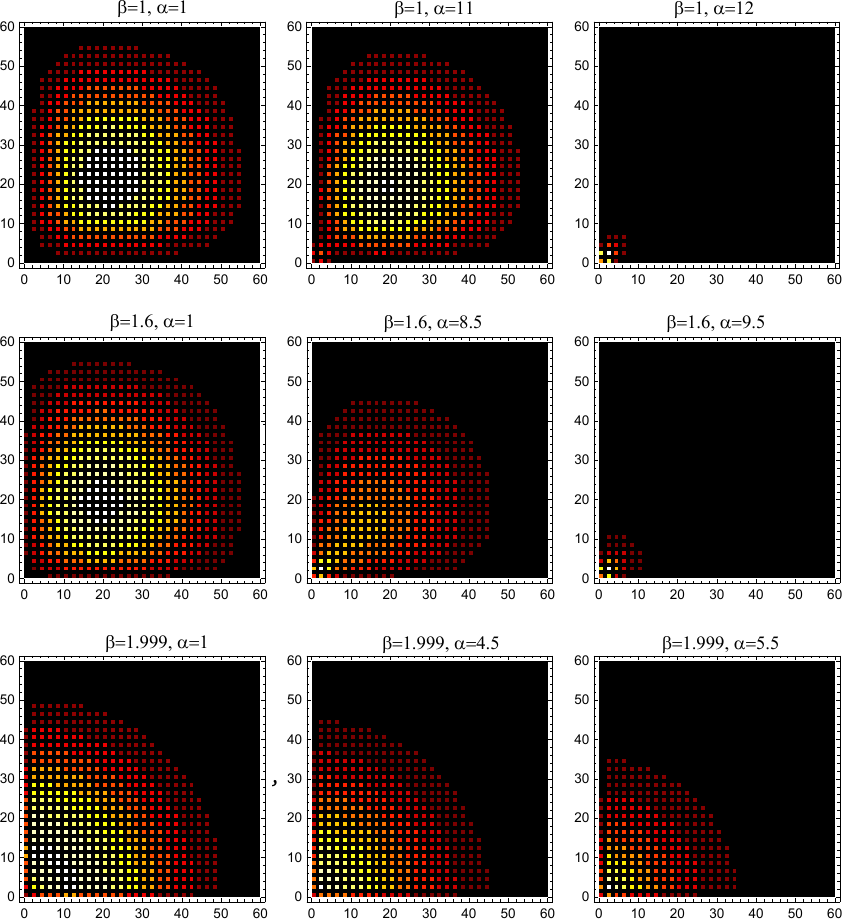}
\caption{Density plots of the end-monomer distributions for $N = 600$-step paths originating at $(0, 0)$ and confined to the first quadrant of a lattice of size $L \times L = 60 \times 60$, shown for various Boltzmann weights at the boundaries ($\beta$) and at the corner ($\alpha$). The shrinking of the distribution indicates localization.}
\label{fig:02}
\end{figure}

Even the numerical computations based on the exact transfer-matrix approach do not permit to make a definite statement. We have performed computations for $N=500$--step paths on the square lattice $L\times L$ with $L=60$, and obtained the snapshots of the last monomer density distribution in the first quadrant at different values of $\beta$ and $\alpha$ for ensemble of paths starting at the point $(0,0)$. These density plots are shown in \fig{fig:02}. The shrinking of the distribution indicates the localization transition. The values $\alpha_{cr}(\beta))$ computed far from $\beta_{cr}=2$ are in a very good agreement with the critical values presented in Table \ref{tab:03}. However at $\beta\to\beta_{cr}$ the numerical results become less and less certain. 

Let us note that the polymer adsorption in the wedge with different weights at the boundaries and the corner has been qualitatively and numerically considered in the work \cite{kardar}. Our results provide the solid background for analytic computations of the corresponding transition.

\subsection{Annealed vs quenched localization: criteria for coincidence and distinction}
\label{sect:4f}

In Section \ref{sect:4a} we pointed out that the relation between annealed and quenched localization transitions depends on which critical threshold is reached first: if the "boundary localization" threshold, $\beta_{cr}=2$, is reached before the "corner localization" one, $\alpha_{cr}$, then both transitions coincide and occur at $\beta_{cr}$, as in the annealed case; however, if $\alpha_{cr}$ is reached first when $\beta \to \beta_{cr}$, the quenched transition occurs at $\alpha_{cr}$, indicating a distinct behavior due to marginal relevance of the disorder.

Given the dependencies of $\beta$ and $\alpha$ on the parameters of the disorder distributions, as shown in Table \ref{tab:01}, we can continuously vary these parameters and observe which critical value, $\beta_{cr} = 2$ or $\alpha_{cr} \approx 5.05185$, is reached first. The results are summarized below. From the Table \ref{tab:03} we see that at the critical  "boundary localization" weight, $\beta \equiv \beta_{cr}=2$, the corresponding critical "corner localization" weight, $\alpha_{cr}$, takes the value $\alpha_{cr}= 4$. However if we step slightly away from the critical value taking for example $\beta=1.999$, the critical corner weight is already $\alpha_{cr}\approx 5.052$ which is essentially larger than 4.

At the "boundary localization weight" $\beta=\beta_{cr}=2$ the exact value of the "corner localization weight", $\alpha_{cr}$, in the thermodynamic limit coincides with the squared value of $\beta_{cr}$: $\beta_{cr}=\alpha_{cr}^2$. Since for any parameters of any disorder distribution $Q(u_j)$ the inequality 
$$
\alpha = \left.\la \left(e^{u_j}\right)^2 \ra\right|_{Q(u_j)} \ge \beta^2 =\left(\left. \la e^{u_j} \ra\right|_{Q(u_j)}\right)^2
$$ 
holds, the derived condition $\beta_{\mathrm{cr}} = \alpha_{\mathrm{cr}}^2$ guarantees that, in the thermodynamic limit, the corner weight $\alpha_{cr}$ is reached first for any choice of disorder distribution parameters. Consequently, quenched disorder is relevant, shifting the transition point away from that of the system with annealed disorder.

However for the values of $\beta$ in the vicinity of $\beta_{cr}=2$ there is a room for ambiguities which we have called "the gray zone" since in this region the difference between annealed and quenched transitions depend on the details of the disorder distribution. The corresponding examples are summarized below. Take the value $\beta=1.999$ which according to the Table \ref{tab:03} gives $\alpha_{cr}\approx 5.052$.
\begin{itemize}
\item For Poissonian disorder (see \eq{eq:02}) we determine $\mu$ from the equation $e^{(e-1)\mu} =\beta = 1.999$ and get $\mu \approx 0.4031$ which gives $e^{(e^2-1)\mu} =\alpha^{(P)}_{cr} \approx 13.1373$. Since $\alpha^{(P)}_{cr}>\alpha_{cr} \approx 5.05185$, the transition at the corner in 2D replica plane happens first, signaling the distinction of quenched and annealed transition points.
\item For Asymmetric bimodal disorder (see \eq{eq:03}) we consider two different values of $p$: $p_1=0.5$ and $p_2=0.9$. For  values $u$ we have: $u(p_1)=1.3164$ and $u_{cr}(p_2)=0.7747$, which are the solutions of the equation $p_{1,2}e^{u_{cr}}+(1-p_{1,2})e^{-u_{cr}} =\beta_{cr}\equiv 1.999$. The corresponding values of $\alpha_{cr}$ are: $\alpha^{(B)}_1=6.992$ and $\alpha^{(B)}_2=4.2589$. Since $\alpha^{(B)}_1>5.05185$ and $\alpha^{(B)}_2<5.05185$, the transition for $p_1=0.5$ at the corner in 2D replica plane happens first signaling the relevance of the quenched disorder, while for $p_2=0.9$ the transition at the boundary happens first and the quenched disorder is not relevant meaning the coincidence of transition points in annealed and quenched systems.
\item For Gaussian disorder with zero mean $\nu=0$ we determine $\sigma$ from the equation $e^{\sigma_{cr}^2} =\beta_{cr}\equiv 1.999$ and get $\sigma_{cr} \approx 0.83256$ which gives $e^{4\sigma_{cr}^2}=\alpha^{(G)} \approx 15.968$. Since $\alpha^{(G)}>5.05185$, the transition at the corner in 2D replica plane happens first, meaning that the transitions in annealed and quenched systems do not coincide. However taking $\nu=0.6$, we get according to our prescription $\alpha^{(G)} \approx 4.8095 < 5.05185$ making the quenched disorder irrelevant.
\end{itemize}

\section{Discussion}
\label{sect:5}

In this work, we studied the localization transition of a (1+1)-dimensional directed random walk interacting with a corrugated impenetrable substrate, modeled as a quenched, site-dependent random potential. Using the exact computations of the disorder-averaged first and second moments of the partition function, we investigated whether the critical points for annealed and quenched pinning transitions coincide or differ.

Our analysis shows that in the thermodynamic limit quenching of the potential is always relevant and the transition points in systems with quenched and annealed disorders do not coincide. However in the vicinity of the critical value $\beta_{cr}=2$ (in the "gray zone") coincidence or distinction of transition points is hard to determine numerically and the result depends on the type of disorder distribution. 

The key mechanism is the competition between two different localization regimes in the two-replica system: adsorption at the boundary, governed by the parameter $\beta = \langle e^{u_j} \rangle$, and adsorption at the corner, governed by $\alpha = \langle e^{2u_j} \rangle$. By solving the two-replica model exactly and identifying the critical values $\beta_{cr}=2$ for annealed system, and $\alpha_{cr}(\beta)$ for quenched system (see Table \ref{tab:03}) , we established a clear criterion: if under the variation of parameters of the disorder starting from the delocalized phase, the value $\beta_{cr}$ is reached first, the annealed and quenched transitions coincide; however if the value $\alpha_{cr}(\beta)$ is reached first, they differ, indicating the marginal relevance of quenched disorder.

Applying this criterion to several disorder types in the gray zone, we found that Poissonian disorder, as well as  Gaussian disorder with zero mean always lead to clearly distinguishable transition points, confirming quenched disorder relevance. In contrast, for asymmetric bimodal disorder and for Gaussian disorder with a nonzero mean, the outcome depends on the asymmetry parameters $p$ and $\nu$. This result may reconcile conflicting viewpoints, providing a framework for identifying when transition is sensitive to the type of the disorder.

\begin{acknowledgments}
We are grateful to the organizers of the Integrable Systems Group seminar at BIMSA, where this work was initiated. SN thanks D. Gangardt, M. Tamm, and O. Vasilyev for numerous discussions on this topic over the course of several years. We also acknowledge the insightful comments of J. M. Luck, which helped us to formulate our conclusions more clearly.
\end{acknowledgments}

\begin{appendix}

\section{Asymptotic behavior of $Q(0,0)$ for $\beta\geq2$}
\label{app:1}

Despite our main concern is the region $1<\beta<2$, here, for completeness, we consider the case $\beta>2$. Since we want to extract $[x^0]$ term of \eqref{eq:24}, we should require $(\beta-1)x^2<1$, otherwise the coefficient of $R_1(x)$ is not expanded as a formal series in $x$. Thus, instead of the unit circle, we take for integration the curve $|C|=\frac{1}{\sqrt{\beta-1}}-\epsilon$. 

The function $F(0,0)$ is still given by equation \eqref{eq:31}. The difference is that we do not take the thermodynamic limit $t \to \tfrac{1}{4}$, but instead consider the limit $t \to \tfrac{\sqrt{\beta - 1}}{2\beta}$. Formally speaking, the branch cut inside the contour must not cross the circle $C$; in the limiting case, it is allowed to touch the circle only at the endpoints of the cut. In this case, we have:
\be
\left.\sqrt{x^2-4 \left(-t x^2-t\right)^2}\right|_{x=\pm\frac{1}{\sqrt{\beta-1}}}=0
\ee
This equation has a solution $t=\frac{\sqrt{\beta-1}}{2\beta}$. By \eqref{eq:36}, in the limiting case, the residue part is zero, and all contributions come from the branch-cut part:
\be
F(0,0)=-\frac{1}{2\pi i}\oint_C\frac{(x-1) (x+1) \left(\beta  \sqrt{x^2-4 \left(-t x^2-t\right)^2}\right)}{2 \left(\beta  x^2-x^2-1\right) \left(\beta ^2 t^2+\beta ^2 t^2 x^4+2 \beta ^2 t^2 x^2-\beta  x^2+x^2\right)}dx
\label{eq:40}
\ee
Eq.\eqref{eq:40} is a function of $x^2$ and has two poles at $\pm \frac{1}{\sqrt{\beta-1}}$. They coincide with the ends of the branch cut and $F(0,0)$ diverges. Thus for arbitrary $\alpha>\beta^2$ the roots of \eqref{eq:29} always exists and at $\beta\geq 2$ there is only one localized phase such that the walk is trapped near the origin $(0,0)$ in the 2D plane. This can be understood in terms of the results of Section \ref{sect:4d}: (i) for $\beta > 2$, the function $F(0,0)$ is characterized by poles but not branch cut singularities; while (ii) for $\beta=2$, by \eqref{eq:30}, one has 
$$
[t^{2j}]F(0,0)\sim \left([t^{2j}]\left(\frac{1}{\sqrt{1-4t^2}}\right)\right)^2
$$ 
and the partition function still diverges.

Let us compute $F(0,0)$ by expanding it as a formal series of $t$ in first turn, as a formal series of $1/x$ in second turn, and finally collect the $[x^0]$--degree terms. In the Table \ref{tab:4} we compare the numerical values of $F(0,0)$ for two different precisions in $t$, $O^{20}(t)$ and $O^{25}(t)$, in the limit $t\to \frac{\sqrt{\beta-1}}{2\beta}$ for two different values of $\beta$:
\begin{table}[ht]
\centering
\begin{tabular}{cc}
\hline \hline
~~Value of $\beta$~~ & ~~$F(0,0)$ with precision $O^{20}(t)$ \\ \hline
$\beta=2.0$ & $F(0,0)=1.82239$ \\ \hline
\rowcolor{gray!20} $\beta=4.0$ & $F(0,0)=31.2881$ \\ \hline \hline
\end{tabular} \qquad
\begin{tabular}{cc}
\hline \hline
~~Value of $\beta$~~ & ~~$F(0,0)$ with precision $O^{25}(t)$ \\ \hline
$\beta=2.0$ & $F(0,0)=1.87666$ \\ \hline
\rowcolor{gray!20} $\beta=4.0$ & $F(0,0)=55.8692$ \\ \hline \hline
\end{tabular}
\caption{Divergence of $F(0,0)$ in the localized phase ($\beta>2$) for two different precisions in $t$: Left panel -- precision $O^{20}(t)$; Right panel -- precision $O^{25}(t)$.}
\label{tab:4}
\end{table}

The function $F(0,0)$ diverges for any $\beta \geq 2$. This is most evident at $\beta = 4.0$ (see Table \ref{tab:4}), where the value of $F(0,0)$ computed with precision $O^{25}(t)$ is nearly twice as large as the value computed with precision $O^{20}(t)$. For smaller values of $\beta$ within the range $\beta \geq 2$, the divergence is harder to detect clearly and requires higher numerical precision, which is computationally challenging.

\end{appendix}

\bibliography{bibliography}

\end{document}